\documentclass[prd,preprint]{revtex4}

\usepackage{graphicx}
\usepackage{dcolumn}
\usepackage{bm}

\begin{document}
\title{Current Short--Range Tests of the\\
Gravitational Inverse Square Law}
\author{Joshua C. Long}
\affiliation{Los Alamos Neutron Science Center, LANSCE-3, MS-H855, 
Los Alamos NM 87545}
\author{John C. Price}
\affiliation{Department of Physics, University of Colorado, Boulder CO
80309}
\date{\today}

\begin{abstract}{Motivated in large part by the possibility of observing
signatures of compact extra dimensions, experimental searches for
deviations from Newtonian gravity at short distances
have improved in sensitivity by many orders of magnitude in the past 
five years.  We review the essential features of the experiments
responsible for the current limits on new effects in the range from a 
few microns to a few centimeters, and discuss prospects for the near future.}
\end{abstract}


\maketitle

\section{Introduction}
There are several motivations for the experimental study of the
gravitational force at short distances.  Most generally, gravity is
poorly understood below one centimeter.  Up until the year
2000 it was not known if gravity obeyed the inverse square law below 
distance scales as large as several millimeters.  Second (and
closely related) is the opportunity to explore a large parameter space for new
forces in nature.  Even with the recent advances in short--distance
measurements, experimental limits allow for new forces at least a 
million times stronger than gravity acting over distances resolvable 
to the unaided eye.  Third, due to developments in theoretical 
particle physics,  short--range gravity experiments have become 
laboratories to test for specific new physics beyond the 
Standard Model.  Many recent models attempting to unify 
gravity and the other fundamental forces in the same theoretical 
framework predict modifications of gravity in the range of about a
millimeter.  Among these, predicted signatures of extra spatial
dimensions have been most responsible for stimulating new experiments.

In these experiments, new phenomena, including the influence of
extra dimensions, are expected to manifest themselves as 
departures from the Newtonian inverse square law.  A common way to 
parameterize these effects is with the Yukawa
interaction.  The potential energy due to gravity and an additional Yukawa
force between test masses $m_{1}$ and $m_{2}$ is given by:  
\begin{equation}
V = -\frac{G m_{1}m_{2}}{r_{12}}[1+\alpha\exp(-r_{12}/\lambda)],
\label{eq:yukawa}
\end{equation}
where $G$ is the Newtonian gravitational constant, $r_{12}$ is the
distance between the test masses, $\alpha$ is the strength of the new
interaction relative to gravity, and $\lambda$ is the range.  The
Yukawa potential arises when the 
interaction between test masses $m_{1}$ and $m_{2}$ in the above 
expression is mediated by a massive field.  In the case of a massless 
mediator with infinite range, such as the graviton, the usual $1/r$
potential is recovered.
For cases in which the mediator couples to some quantity other than
mass (for example, baryon number), $\alpha$ can acquire a dependence
on the composition of the test masses.  This effect is expected to be
at most a small fraction of $\alpha$.  It is addressed in experimental
tests of the equivalence principle, which are conducted over ranges in
which gravity is relatively well--measured.  

This review concentrates on the search for new effects in the
interaction range between a few microns and a few centimeters.  As the
experiments in this range are at or near gravitational
sensitivity, it is here where phenomena associated with the extradimensional 
models of greatest interest are most likely to appear in
the near future.  There are also other specific predictions of new
physics in this regime, some indirectly related to the physics of
extra dimensions, which may soon be observable.  Experimental programs
motivated in part by the physics of extra dimensions have begun to
explore other distance ranges as well.  A significant effort is active
in the range below 100~nm in connection with recent Casimir force
measurements, as reviewed in Refs.~\cite{fischbach} and~\cite{bordag_pr}.

\begin{figure}[htbp]
\begin{center}
\includegraphics[width=10cm]{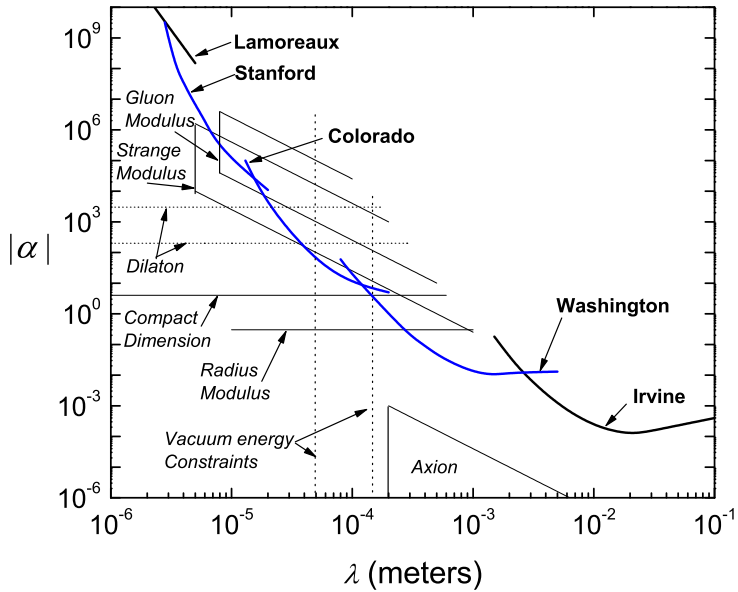}
\caption{\label{fig:limits} Parameter space for deviations from
  Newtonian gravity in which the strength $\alpha$ of a hypothetical
  new Yukawa--type force is plotted versus the range $\lambda$. The
  current experimentally excluded region is above and to the right of
  the solid, bold curves.  Curves in black are published results;
  curves in blue are unpublished but available in preprints. (A
  slightly weaker limit from the Washington group has been
  published~\cite{washington_prl}.)  Curves 
  labeled Stanford and Irvine are 1$\sigma$ limits; others are
  2$\sigma$ limits.  Theoretical predictions are represented by
  the fine and dashed lines.  For the moduli, dilaton, and compact
  dimension theories, the upper bounds on $\lambda$ of the regions
  shown are set at the approximate experimental limits known at the time the
  theories were proposed; the same is true of the $\alpha$ upper
  bounds for the vacuum energy prediction.\\
  {\it Note}: As this work was being submitted, the
  authors were informed of the publication of the Colorado curve
  above~\cite{colorado_nature}.}
\end{center}
\end{figure}

To date, no convincing signal of a departure from Newtonian gravity
has been observed. In the absence of signals, experiments commonly 
report results in terms of limits on the
strength parameter $\alpha$ for a given range $\lambda$.  The current
experimental limits are shown in Fig.~\ref{fig:limits}, together with
recent theoretical predictions. 
\section{Theoretical Motivations}
Several predictions of new physics which can be tested in
current short--range experiments arise in models addressing the
unification of the fundamental forces.  The leading candidate models
are inspired by string or M-theory, which has to be formulated in more
than three spatial dimensions.  The extra dimensions are typically
compactified at a scale on the order of the Planck length 
($L_{P} \sim 10^{-35}$~m or $L^{-1}_{P} = M_{P} \sim 10^{19}$~GeV), 
the presumed gravitational unification scale, or concealed from direct
detection by other mechanisms.  Recently several
models have been discovered which hinge on at least one of the extra
dimensions remaining larger than $L_{P}$, with significant consequences for
short--range experiments.  String theories also contain scalar
fields called moduli, which parameterize the size and shape of the extra
dimensions and can mediate new short-range forces.  Theoretical models are 
treated extensively in other parts of this volume; in this section 
we provide a brief description of some of the predictions.

In the recent model proposed by Arkani-Hamed, Dimopoulos, and Dvali
(ADD), unification can occur at a scale $L^{-1}=M^{*} << M_{P}$,
with the relative weakness of the gravitational interaction arising as a 
consequence of $n$ ``large'' extra dimensions in which
gravitons propagate but the Standard Model fields do not~\cite {ADD}. 
The size $R$ of the extra dimensions is related to the unification scale
and the usual Planck scale by:
\begin{equation}
R^{n}=M_{P}^{2}/M^{*(2+n)}
\end{equation}
Setting $M^{*} = 1$~TeV, and effectively eliminating the hierarchy
problem (the vast discrepancy or ``desert'' between the weak and
Planck scales), the choice of $n = 1$ implies 
$R \sim 10^{13}$~m, clearly ruled 
out by astrophysics.  However, the choice of $n = 2$ 
implies $R \sim 1$~mm, with the consequence that the gravitational
force will behave according to a $1/r^{4}$ law below this scale.
As the scale $R$ is approached from above, Yukawa 
corrections are predicted; the model illustrated in
Fig.~\ref{fig:limits} predicts $\alpha = 4$~\cite{floratos,kehagias}. 
In the absence of signals for these large compact dimensions,
short--range gravity experiments can set limits on the unification
scale $M^{*}$, which is also probed directly in collider experiments.

A consequence of eliminating the hierarchy problem in favor of
embedding the Standard Model on a 3-dimensional brane in a higher
dimensional bulk is that the success of the ``desert'' in
explaining the approximate symmetries responsible for a variety of
phenomena (e.g., small neutrino masses and proton stability) no longer
applies.  In response to these issues, two authors of the ADD model consider a
scenario in which the appropriate symmetry--breaking processes occur
at a higher order on distant branes, with suppressed
information about this breaking transmitted to the Standard Model
brane via messenger fields~\cite{arkani}.  The messenger fields should
be easily detectable by a short--range macroscopic force search if
they have mass of $10^{-3}$~eV or less, as they are predicted to
have couplings on the order of $\alpha \sim 10^{6}$. 

Several predictions arise in supersymmetric models with 
extra dimensions compactified on the order of the weak scale.  
Antoniadis, Dimopoulos and Dvali consider a model in which
supersymmetry is broken at $\sim 1$~TeV by a compactification
process, and show that the modulus associated with the 
dimension left at this scale acquires mass in the sub--eV range and a
coupling of $\alpha \sim 1/3$~\cite{antoniadis}.  The prediction for
this ``radius'' modulus is shown in Fig.~\ref{fig:limits}.  
Considering the problem of the stabilization of the extra dimensions 
in related models, Chacko and
Perazzi predict a scalar field or ``radion'' corresponding to the 
size fluctuations of TeV--sized extra dimensions with a coupling of 
$\alpha \sim 30$ and a range $\lambda \sim 40~\mu$m~\cite{chacko}.  

The other moduli predictions in Fig.~\ref{fig:limits} derive 
from an earlier model in which supersymmetry is broken near 100~TeV by a
gauge--mediated process.  This leads to moduli with masses in the 
sub--eV range and couplings as strong as 
$\alpha \sim 10^{6}$~\cite{dimopoulos}.

The dilaton is another scalar in string theory whose vacuum expectation
value determines the string coupling constant.  Dilaton mass can also arise
from supersymmetry breaking but is generally sensitive to unknown
physics.  Predictions for the dilaton coupling vary widely in
the literature; the lower and upper bounds in Fig.~\ref{fig:limits}
are representative~\cite{taylor,ellis,kaplan}. 

Fig.~\ref{fig:limits} also shows two predictions of new macroscopic 
effects not directly related to string theory.  The axion, a light
pseudoscalar motivated by the strong CP problem of QCD, is predicted
to mediate forces between unpolarized test masses~\cite{moody}.  The allowed
range between 200~$\mu$m and 20~cm in Fig.~\ref{fig:limits} 
is constrained by laboratory and astrophysical observations~\cite{rosenberg}, 
and the coupling by measurements of the neutron electric dipole moment.
The remaining prediction is motivated by the cosmological constant
($\Lambda$) problem, the discrepancy between the observed flatness 
of the universe and the extreme curvature expected from the vacuum energy
contributions of the Standard Model fields.  Several models
\cite{sundrum,beane,schmidhuber} have attempted to address this problem by
introducing new interacting quanta
with a range on the order of $\Lambda^{-1/4}\sim 100~\mu$m. 
\section{Experimental Challenges}
The principal limitation to testing gravity at short distances is the
scaling of the signal force with the size of the apparatus.  If all
dimensions of a gravitational experiment are scaled by the same factor
$r$, gravitational attraction behaves as $r^{4}$, so that signal
forces become extremely weak at small distances.  For example, 
expressing the gravitational force between two spherical test objects 
(radius $\sim$ separation $\sim r$) in terms of their densities $\rho$
yields: $F \sim G \rho_{1} \rho_{2}r^{4}$.  For typical test mass densities
($\rho \sim 20~\mbox{g/cm}^{3}$), substituting $r = 10$~cm 
gives $F \sim 10^{-5}$~N.  Measuring a
force of this magnitude was the challenge met by Cavendish with his
torsion balance experiment, published in 1798.  In order to measure gravity 
at 100~$\mu$m, the experimenter is faced with scaling the size of the 
test masses to this range as well, so as not to be overwhelmed by the 
attraction of the additional mass at
larger scales.  Setting $r = 100~\mu$m yields $F \sim 10^{-17}$~N,
which explains in large part why it has taken two centuries to achieve
measurements sensitive to gravitational effects below a millimeter.

At the same time, backgrounds due to known physics increase rapidly
at short distances.  Electrostatic effects due to surface potentials
increase as $r^{-2}$, and magnetic forces due to contaminants increase 
as $r^{-4}$.  At 
the lower end of the range considered ($\sim 1~\mu$m), the Casimir 
force (of experimental interest itself) also 
becomes visible and quickly begins to dominate.  
The following sections are devoted to a description of current
short--range experiments, with emphasis on the strategies
used to overcome the dual challenge of extremely small signals and
rapidly increasing backgrounds.

\section{Classical Gravitation Experiments}
\subsection*{\it The Irvine 2--5~cm Experiment and Cryogenic Torsion
  Pendulum}
The highly linear and extremely sensitive torque response of thin filaments
under tension has made the torsion balance the instrument 
of choice for measuring weak forces at laboratory scales.  
Between 3~mm and 3~cm, the best constraints on new forces derive
from the 2--5~cm experiment in the group of R. Newman at the
University of California at Irvine~\cite{irvine}.  In this experiment,
a cm--sized 
cylindrical test mass was suspended from one end of a torsion balance 
inside a long massive tube 
(a null geometry for $1/r^{2}$ forces).  As the tube was translated
periodically perpendicular to its axis, no deflection of the balance
above that associated with a small Newtonian edge effect was
observed.  Torque sensitivities of about $10^{-14}$~Nm were obtained,
yielding the limit curve in Fig.~\ref{fig:limits}. 

A principal limitation to the sensitivity of this experiment was
the effect of tilt.  Motion of the tube along its guide
track caused a slight tilt in the large block supporting the balance and
a resulting phase--correlated motion of the balance suspension point.
While the tilt was largely compensated with external instrumentation,
the residual effect remained one of the main sources of systematic
error.  Magnetic
effects presented another large source of systematic uncertainty,
while torque measurement sensitivity was limited by a combination of
instrument noise, seismic noise, and spurious
gravitational forces not associated with the motion of the tube.  

\begin{figure}
\begin{center}
\includegraphics[width=10cm]{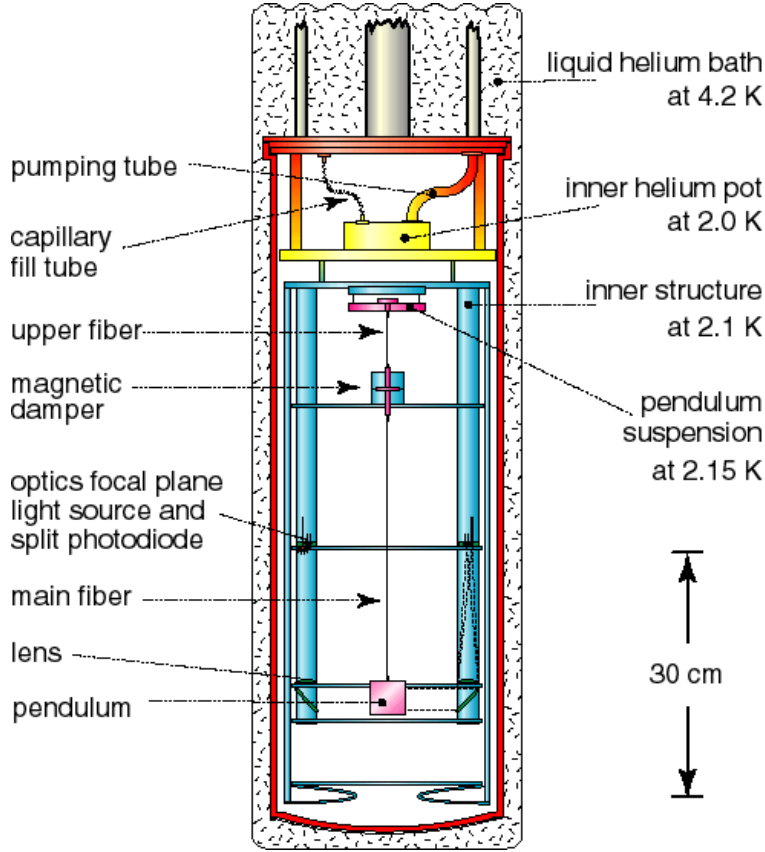}
\caption{\label{fig:irvine} Schematic of the Irvine--Washington
 cryogenic torsion pendulum.  Source masses are suspended alongside
 the cryostat at the level of the pendulum mass.  Source:
 http://www.physics.uci.edu/gravity/Gcryo.htm}
\end{center}
\end{figure}

For the past several years, the Irvine group has been constructing a
torsion pendulum which operates at cryogenic temperatures ($T\sim
2$~K), in collaboration with the group of P. Boynton at the University of
Washington (Fig.~\ref{fig:irvine}).  This approach has several 
advantages~\cite{cryo}, of which we mention only a few.  For one, the 
pendulum may be operated in 
the frequency mode, which is highly insensitive to the effect of tilt.  In 
this method, pioneered 
by the Boynton group~\cite{Boynton}, the resonant frequency of the pendulum is
monitored for the shifts induced by the force gradient generated by 
an external source mass.  At room temperature, the temperature 
dependence of the torsion fiber elastic properties limits the usefulness of 
this mode, but at cryogenic temperatures this dependence is small and 
temperature stability is easy to maintain.  Operation at cryogenic 
temperature also allows for the use of superconducting shielding to
reduce magnetic backgrounds. 

Finally, cryogenic operation can reduce instrument 
noise considerably, inasmuch as this effect is dominated by thermal
noise due to dissipation in the force--sensitive detector.  In the 
thermal noise limit, the sensitivity of this experiment 
(and all others discussed below) to new forces is given by: 
\begin{equation}
\alpha \sim \frac{1}{\rho_{1}\rho_{2}}\sqrt{\frac{\omega T}{Q\tau}},
\label{eq:sensitivity}
\end{equation}
where $\rho_{1}$ and $\rho_{2}$ are the test mass densities, $\omega$
is the operational frequency of the experiment, $T$ is
the temperature, $Q$ is the detector mechanical quality factor, and
$\tau$ is the experimental integration time.  In addition to the
improvement in sensitivity expected from a reduction in $T$ from 300~K to
2~K, further improvement can be expected from a gain in $Q$ which is 
observed in many materials at low temperatures.

Construction of the cryogenic experiment is complete and a precision
measurement of the Newtonian constant $G$ is in progress.  A test of 
the inverse square law is planned which will have maximum sensitivity
for forces with a range of about 15~cm.   For the latter experiment,
additional sensitivity is expected from a re-optimized test mass
design based on a harmonic expansion of the mass and field multipole
moments.  In a method similar to that used in previous equivalence
principle tests, the pendulum and source mass are configured to
suppress Newtonian mass and field moments, respectively, so that
classical gravitational effects enter into the measurement only to second
order or higher~\cite{moore}.  At the same time, non--Newtonian
moments can be enhanced. If all noise sources can be reduced to near
the level of the thermal noise, an improvement in sensitivity of about 
two orders of magnitude over previous results is expected,
corresponding to the projected limit shown in Fig.~\ref{fig:proj}.

\subsection*{\it The E\"{o}t-Wash Short-Range Experiment}
\begin{figure}
\begin{center}
\includegraphics[width=10cm]{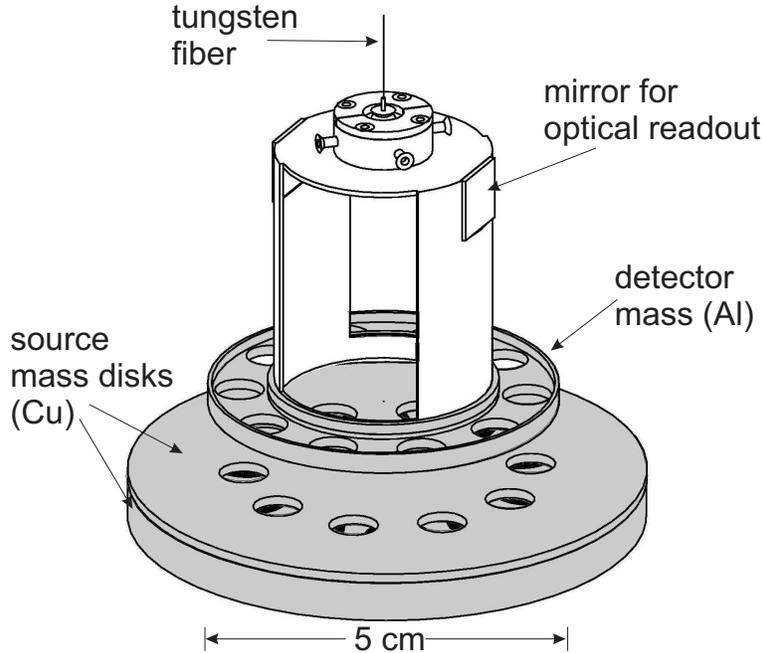}
\caption{\label{fig:eot-wash} E\"{o}t-Wash short--range torsion
  pendulum experiment. Source: Ref.~\cite{washington_prl}}
\end{center}
\end{figure}

Between $100~\mu$m and 3~mm, the most sensitive limit on new forces 
has been attained by the short--range torsion pendulum experiment in the 
E\"{o}t-Wash group at the University of Washington
(Fig.~\ref{fig:eot-wash})~\cite{washington_prl}.  The force--sensitive part of 
the torsion pendulum consists of a 1~mm thick
aluminum annulus with an array of ten equally spaced holes.  The source
mass consists of a stack of two copper disks, each a few
millimeters thick, with similar arrays of holes.  The source rotates 
approximately once every two hours, torquing the pendulum 10 times
per revolution.  This arrangement has several clever features.
For one, the signal occurs at 10 times the revolution rate, allowing 
easy discrimination from vibrations associated with the source
drive.   Also, in the source stack, the angular position of the lower
disk is offset relative to the upper disk so as to cancel the torque 
associated with the long--range Newtonian attraction between the test 
masses, while leaving possible sub--millimeter forces unaffected. 

Not shown in Fig.~\ref{fig:eot-wash}, an essential component of the
experiment is an electrostatic shield.  This is made from a $20~\mu$m
thick beryllium--copper foil stretched on a frame between the 
source and pendulum.

The torque as a function of test mass separation was measured for two
prototypes of this apparatus.  Torque sensitivities of about $10^{-16}$~Nm were
attained, and no deviation from the expected
Newtonian signal was observed for test mass separations down to about
$200~\mu$m.  From these data, the preliminary results shown 
in Fig.~\ref{fig:limits} were obtained~\cite{washington_xxx}.  
This is the shortest--ranged experiment to date to have attained 
gravitational sensitivity.

Improvements to this experiment currently under way include 
thinner pendulum and source disks, both constructed of copper or molybdenum
for greater test mass density.  The hole arrangements have been further
optimized for Yukawa signals and cancellation of Newtonian gravity.
Better vibration isolation and a thinner shield should significantly
reduce the test mass separation.  The projected limits expected 
with this technique are shown in Fig.~\ref{fig:proj}.

\section{High Frequency Techniques}
With low operating frequencies and high mechanical quality factors
($\sim~10^{5}$), torsion balance experiments are very
attractive from the point of view of thermal noise, even at room
temperature.  However, their sensitivities are usually limited by
other backgrounds, including low--frequency vibrations which can also
limit the minimum practical test mass separation.  
Recently, several experiments have been developed using
high--frequency techniques, which show promise for operation at the
thermal noise limit and for attaining smaller test mass separation.

\subsection*{\it Colorado Torsional Oscillator Experiment}
\begin{figure}
\begin{center}
\includegraphics[width=10cm]{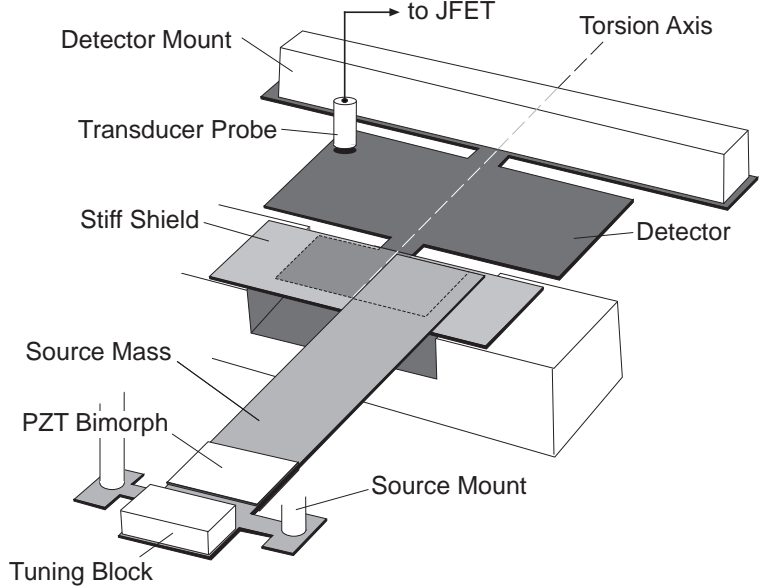}
\caption{\label{fig:colorado} Central components of Colorado
  experiment. Figure is to scale. The source mass has approximate
  dimensions 7~mm $\times$ 40~mm $\times$ 0.25~mm and is driven at
  the detector
  resonance by a PZT (lead zirconate titanate) bimorph. Detector
  motion is monitored with a capacitive probe suspended above a rear
  corner and connected to a JFET amplifier.}
\end{center}
\end{figure}

An approach for improving the experimental sensitivity at
$100~\mu$m and below has been pursued for the past several years by
the authors and their collaborators at 
the University of Colorado~\cite{colorado_npb,colorado_mg9}.  The
basic design is shown in Fig.~\ref{fig:colorado}.  Planar test 
mass geometry is used in order to concentrate as much of the available 
test mass
density as possible at the range of interest.  The flat plates are 
nominally a null geometry 
with respect to
$1/r^{2}$ forces, which is important for suppressing Newtonian forces
relative to new short--range effects.  
The test masses are fabricated from tungsten 
wafers, approximately 250 microns thick.  The source mass consists of
a 4 cm long cantilever,
which is driven mechanically at a natural resonant frequency of the 
detector mass, nominally 1~kHz.  The detector consists of two coplanar 
rectangles joined along 
their central axes by a short segment.  In the resonant mode of
interest, the detector rectangles counter--rotate in a torsional mode
about the axis defined by the segment.

Electrostatic and acoustic backgrounds are suppressed
with a stiff conducting shield, consisting of a gold--plated sapphire wafer,
between the test masses.  The shield currently limits the minimum test
mass separation to approximately 100 microns.

Operation of the experiment at the detector resonant frequency, while
increasing the experimental sensitivity, places strict requirements on
vibration isolation.  At the operational frequency of 1~kHz it is
possible to construct a simple passive vibration isolation system, in
which the test masses and shield are each
suspended from a series of brass disks connected by fine wires
under tension.  This system provides about 200 dB of attenuation at 
the operational frequency~\cite{chan}.  Modular design, visibility, and
easy control of test mass positioning and electrical connections 
allow for rapid characterization of observed signals {\it in situ}.  

Electromagnetic, acoustic and vibrational backgrounds have been
sufficiently suppressed such that the experiment is currently limited
by detector thermal noise.  To reduce the thermal
background, the detector is annealed at 1300 C for
several hours before installation in the experiment, which 
significantly increases its mechanical quality factor.  

In a recent series of runs, no signal above thermal noise was observed
after 22 hours of integration time.  This corresponded to a force 
sensitivity of about $10^{-15}$~N, from which the limit derived in 
Fig.~\ref{fig:limits} is obtained~\cite{colorado_xxx}. 
Development of a stretched--membrane shield and flatter test masses
with higher Q is in progress.  In the absence of backgrounds, the
limits attainable with the current sensitivity of the experiment
(at 300~K) at 
half the present minimum test mass separation are shown in 
Fig.~\ref{fig:proj}.  A cryogenic version of this experiment, with a 4~K
operating temperature and an expected detector $Q$ enhancement of an
order of magnitude, could improve the sensitivity by an additional
factor of 40.

\subsection*{\it Stanford Micro-cantilever Experiment}
\begin{figure}
\begin{center}
\includegraphics[width=10cm]{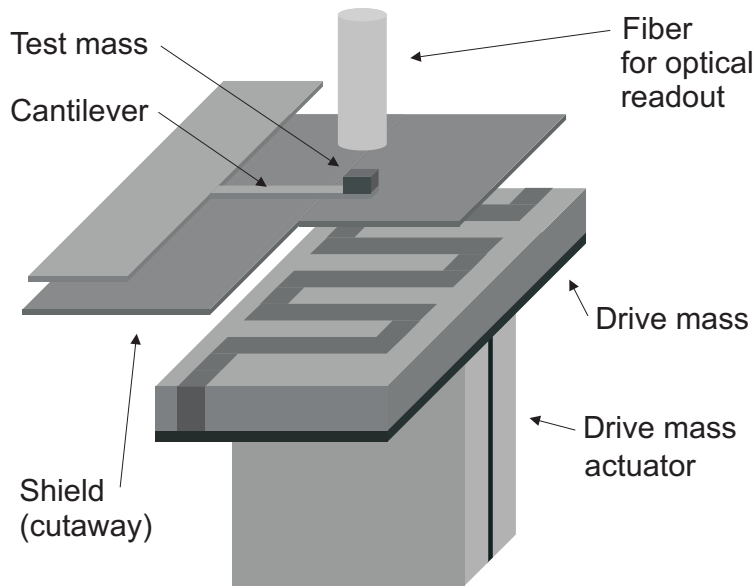}
\caption{\label{fig:stanford} Central components of the micro-cantilever
  experiment at Stanford. Figure not to scale; test mass in actual
  apparatus is approximately a 50~$\mu$m cube suspended 15~$\mu$m
  above shield. Source: Chiaverini J., ``Small Force Detection Using
  Microcantilevers: Search for Sub-millimeter Range Deviation from
  Newtonian Gravity,'' Doctoral Thesis, Stanford University, 2002
  (unpublished).}
\end{center}
\end{figure}

Another high--frequency measurement is in progress in
the group of A. Kapitulnik at Stanford.  This experiment uses
techniques of scanning force microscopy and microfabrication to probe distance
scales on the order of a few tens of microns (Fig.~\ref{fig:stanford}).
The force--sensitive detector mass consists of a $50~\mu$m gold cube
mounted on the end of a silicon nitride cantilever $250~\mu$m long and
less than $1~\mu$m thick. 

The source mass consists of a parallel array of alternating strips of
gold and silicon, each $100~\mu$m wide.  It is scanned
in the horizontal plane below the detector in the direction
perpendicular to the strips, with the resulting periodic variations in density
providing the driving mass--coupled force.  This design offers an 
advantage similar to that of the E\"{o}t-Wash torsion pendulum, in that
the frequency of the source drive mechanism can be set much lower than
that of the expected signal.  In the Stanford
experiment, the mass modulation frequency is matched to the detector 
cantilever resonance (about 300 Hz), resulting in less of
a burden on vibration isolation than for gap--modulated experiments.

An electrostatic shield, fabricated from a silicon nitride 
membrane $3~\mu$m thick with a 100~nm gold plating, is placed between 
the test masses.  It is attached to the base of the cantilever mount, 
leaving a gap of $15~\mu$m below the detector mass.

The limiting background to
this measurement is expected to be thermal noise of the 
force--sensitive cantilever.  To reduce this background, the
experiment is inserted into a liquid helium
dewar for operation at $T\sim 10$~K, where
the expected force sensitivity is about $10^{-17}$~N for integration
times of a few hours.

After 20 minutes of integration time during a run in 2002, data from
this experiment showed evidence of a signal approximately three times
greater than the expected thermal noise~\cite{stanford}.  The signal 
appeared most 
likely to be an electrostatic background, due to modulation of the 
shield caused by surface variations on the source, and a 0.3~V 
potential between the shield and the detector mass.  The phase and
magnitude of the observed signal was not consistent with a
mass--coupled force generated by the source mass, and the data were used 
to obtain the preliminary limits shown in 
Fig.~\ref{fig:limits} representing an upper limit on such forces.
While a projected limit was not available as of this writing,
improvements to the experiment continue, including a 
new source mass with additional conducting and insulating layers 
above the strips to eliminate surface variations.

\subsection*{\it Dusseldorf Torsional Oscillator Experiment}
\begin{figure}
\begin{center}
\includegraphics[width=10cm]{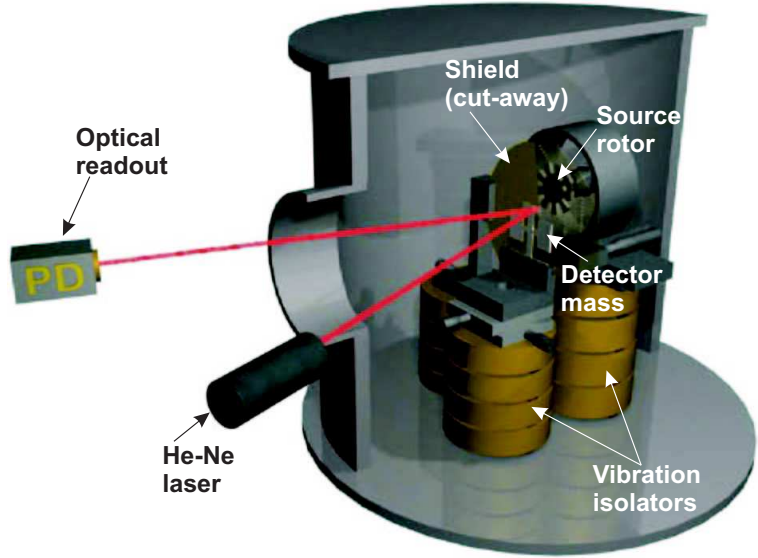}
\caption{\label{fig:dusseldorf} Experiment at Dusseldorf: cut--away
  view of vacuum chamber. Source:
  Haiberger L., L\"{u}mmen N., Schiller S., ``A Resonant Sensor for 
  the Search for Deviations from Newtonian Gravity at Small
  Distances,'' May 2001 (unpublished), also available at\\ 
  http://www.exphy.uni-duesseldorf.de/ResearchInst/WelcomeFP.html}
\end{center}
\end{figure}

An experiment in progress in the group of S. Schiller at the
University of Dusseldorf introduces several new features.  An early
conception is illustrated in Fig.~\ref{fig:dusseldorf}.  Planar
geometry is used, with both test masses mounted vertically.
The detector mass, fabricated from a high--purity silicon wafer, 
is a torsional oscillator similar in design to the Colorado
detector.  Excitation of the source mass at the detector resonance 
(about 5 kHz) is accomplished using a rotating disk with nine
tungsten alloy teeth embedded in a lower--density substrate.

A thin metal plate is placed between the test
masses as an electrostatic shield.  All elements are mounted atop
multi--stage vibration isolation stacks, consisting of viton rings
sandwiched between 1~kg--mass brass disks.

The sensitivity of this experiment is also expected to be limited by
thermal noise.  In the thermal--noise limit, the reduction in
sensitivity associated with the use of a low--density
silicon detector is offset somewhat by the
exceptionally high mechanical quality factors attainable, which are
about an order of magnitude larger than those observed so far in
metal oscillators at room temperature.  Plans for operation of the
experiment at 4~K are under way, where improvement in $Q$ of at 
least two additional orders of magnitude can be expected for silicon
detectors.  The combined effect of the increase in $Q$ with the reduction in
temperature is reflected in the very sensitive projected limits for
this experiment, shown in Fig.~\ref{fig:proj}. 

A room--temperature prototype of the experiment has
been assembled and is under test.  In early runs, a signal
approximately 10 times larger than the detector thermal noise was
observed.  As the signal was present even for large separations of the
test masses, a vibrational background was suspected.  This background
has been reduced with several improvements in the design, including a
more uniform source disk and remote positioning of the drive vibration
isolation stack.  A cryostat for the 4~K version of the experiment 
is also under construction.
\subsection*{\it Padova Casimir Force Experiment}
Below test mass separations of a few microns, the principal background
to macroscopic force searches is expected to be the Casimir effect,
a force which arises due to zero--point fluctuations of the
electromagnetic field.  At the moment,
the best limits on new forces in this range are derived from direct
measurements of this effect, in general by ascertaining the maximum possible
size of a Yukawa--type force for the Casimir plates
still consistent with the experimental residuals.  Below $\lambda =
3~\mu$m, the best published limits are obtained from the
measurement by Lamoreaux in 1997~\cite{bordag_pr,colorado_npb} which used
a torsion balance to measure the Casimir force between a flat 
gold--plated disk and a thin spherical shell~\cite{Lamoreaux}.

\begin{figure}
\begin{center}
\includegraphics[width=10cm]{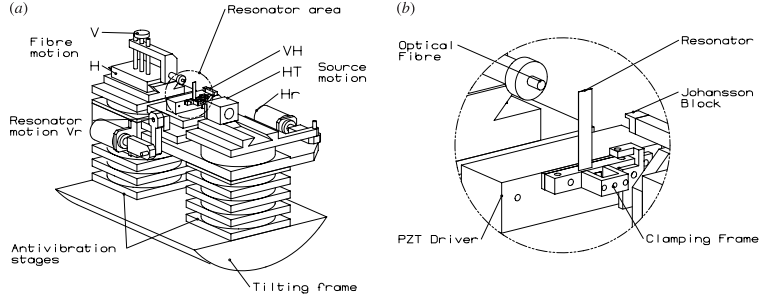}
\caption{\label{fig:padova} (a) Central components and (b) detail of
  the Casimir force experiment at Legnaro laboratories. The source mass is
  mounted on the Johansson block. Source: Ref.~\cite{padova_cqg}.}
\end{center}
\end{figure}

An experiment at Legnaro laboratories by a joint team of the INFN,
Padova and Pavia, and the University of Padova, uses a
frequency--shift technique to measure the
Casimir force between parallel conducting plates 
(Fig.~\ref{fig:padova})~\cite{padova_prl}.  The
force--sensitive detector plate, a $47~\mu$m thick, 2~cm long 
silicon reed with a 50~nm chromium plating, is clamped at its base and 
driven electrically into oscillation at its lowest natural frequency of about
138~Hz.  The source mass, a 5~mm thick, chromium--plated silicon
block, is mounted on a nearby translation stage and brought into very
close proximity to the detector.  The detector resonance is
monitored optically for the frequency shifts induced by static
external force gradients.  

The apparatus is mounted on passive
vibration isolation stacks, and in turn into a large scanning electron
microscope (not shown in Fig.~\ref{fig:padova}) in order to monitor 
the cleanliness and parallelism of the test mass surfaces at the 
micron--sized separations needed for this measurement.
The electrostatic background is due primarily to contact potentials in
the circuit connecting the test masses.  It is eliminated to first
order by nulling the static force between the test masses at small
distances, and the remaining background is subtracted off after
careful fits to the measured force in the presence of applied
potentials at larger gaps.

This experiment was able to obtain a measurement of the Casimir force
to a precision of 15\%, the first unambiguous measurement of this
effect for the parallel--plate geometry.  An earlier version of this 
experiment was
among the first high--frequency measurements used to derive limits on new 
effects~\cite{padova_prd}.  While those limits were not competitive 
with the most stringent results at that time, a re--optimized version
of the present apparatus is expected to improve the limits by at
least an order of magnitude in the range near $1~\mu$m, as shown in
Fig.~\ref{fig:proj}~\cite{padova_cqg}.

\section{Conclusion}
Motivated in large part by suggestions of extra dimensions, three of
the experiments described above have explored approximately 7 square 
decades of the parameter space for new forces in the past five
years~\cite{historical}, and have set significant limits on new 
physics.  The maximum possible size of any extra dimension in the ADD 
scenario has been limited to $200~\mu$m (or $150~\mu$m for $n = 2$ 
equal--sized extra dimensions, corresponding to a lower limit 
on the unification scale of 
$M^{*} \geq 4$~TeV)~\cite{washington_xxx},  a new upper limit on the 
dilaton mass has been set at $8.6 \times 10^{-3}$~eV (assuming a
limit on the coupling of $\alpha \leq 2000$), and much of the 
remaining parameter space for the moduli force scenario of
Ref.~\cite{dimopoulos} has been excluded.
\begin{figure}
\begin{center}
\includegraphics[width=10cm]{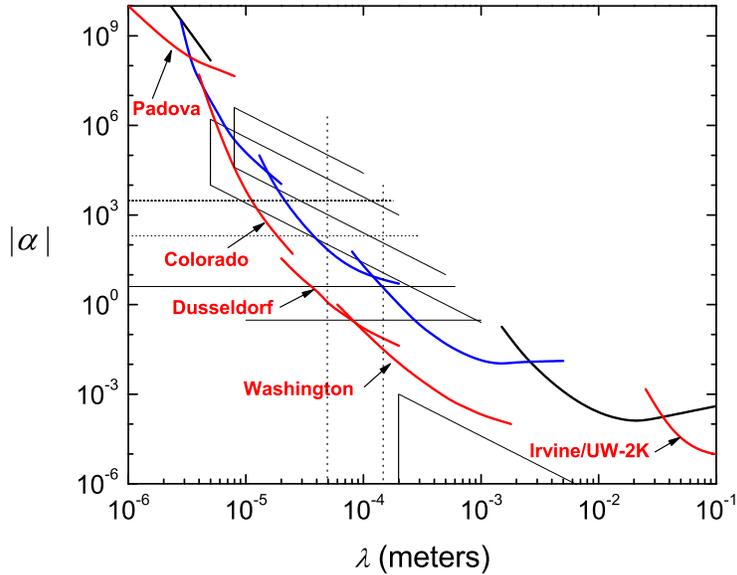}
\caption{\label{fig:proj} The projected limits for several experiments
  currently in progress (red).  The Irvine--UW curve is from M. Moore
  (private communication).  The Washington curve is from 
  http://www.npl.washington.edu/eotwash/shortr.html, and 
  the Dusseldorf curve from Haiberger L., L\"{u}mmen N., Schiller S., 
  ``A Resonant Sensor for the Search for Deviations from Newtonian
  Gravity at Small Distances,'' May 2001 (unpublished).}
\end{center}
\end{figure}

While the torsion pendulum remains the instrument of choice for 
scales above 100 microns, high--frequency experiments show 
great potential for improving the limits at shorter
ranges.  Fig.~\ref{fig:proj} shows the projected limits for the
next few years, considering only some of the experiments currently in
progress.  Experimental coverage of at least 7 additional square
decades of parameter space, with investigation of the region of 
cosmological interest at gravitational strength, seems well within 
reach in the near future.

\section*{Acknowledgements}
We wish to thank R. Newman, P. Boynton, and M. Moore for
description of the cryogenic torsion pendulum experiment, L. Haiberger and 
S. Schiller for materials describing the Dusseldorf experiment, 
and E. Adelberger, J. Chiaverini, C. D. Hoyle, A. Kapitulnik,
R. Onofrio, and G. Ruoso for useful materials and comments.

\end{document}